\documentclass[12pt]{article}
\usepackage{times}
\usepackage{geometry}
\usepackage[utf8]{inputenc}
\usepackage[colorlinks=true,linkcolor=blue,citecolor=blue]{hyperref}
\hypersetup{urlcolor=blue}
\usepackage{enumitem,amssymb}
\usepackage{ragged2e}
\newlist{thematic}{itemize}{8}
\setlist[thematic]{label=$\square$}
\usepackage{pifont}
\usepackage{tabu}
\usepackage{multirow}
\usepackage{longtable}
\usepackage{rotating}
\usepackage{natbib}
\usepackage{wrapfig}

\usepackage{titlesec}
\titlespacing*{\section}{0pt}{.75\baselineskip}{.5\baselineskip}
\titlespacing*{\subsubsection}{0pt}{.75\baselineskip}{.5\baselineskip}
\titlespacing*{\subsection}{0pt}{.75\baselineskip}{.5\baselineskip}

\usepackage[colorinlistoftodos,textsize=footnotesize]{todonotes}
\definecolor{kellygreen}{rgb}{0.3, 0.73, 0.09} 
\definecolor{palegreen}{rgb}{0.6, 0.98, 0.6}

\DeclareRobustCommand{\thinskip}{\hskip 0.16667em\relax}
\def\emdash{---}
\def\d@sh#1#2{\unskip#1\thinskip#2\thinskip\ignorespaces}
\def\Dash{\d@sh\nobreak\emdash}

\newcommand{\ana}{A\&A}

\clearpage
\begin{document}

\thispagestyle{empty}
\onecolumn

\raggedright

\huge
\emph{Astro2020 State of the Profession
Consideration White Paper} \linebreak

\textbf{Realizing the potential of astrostatistics and astroinformatics}

\normalsize

\bigskip

\today

\bigskip

\textbf{Principal Author:}

Name: {\it Gwendolyn Eadie\footnotemark[4]$^,$\footnotemark[5]$^,$\footnotemark[6]$^,$\footnotemark[15]$^,$\footnotemark[17]}	

Email: eadieg@uw.edu
 \linebreak

\textbf{Co-authors:} 

    Thomas Loredo\footnotemark[1]$^,$\footnotemark[19],
    Ashish A. Mahabal\footnotemark[2]$^,$\footnotemark[15]$^,$\footnotemark[16]$^,$\footnotemark[18],
    Aneta Siemiginowska\footnotemark[3]$^,$\footnotemark[15],
    Eric Feigelson\footnotemark[7]$^,$\footnotemark[15],\\
    Eric B. Ford\footnotemark[7]$^,$\footnotemark[15],
    S.G. Djorgovski\footnotemark[2]$^,$\footnotemark[20],
    Matthew Graham\footnotemark[2]$^,$\footnotemark[15]$^,$\footnotemark[16],
    ${\check{\rm Z}}$eljko Ivezi{\'c}\footnotemark[6]$^,$\footnotemark[16],
    Kirk Borne\footnotemark[8]$^,$\footnotemark[15],
    Jessi Cisewski-Kehe\footnotemark[9]$^,$\footnotemark[15]$^,$\footnotemark[17],
    J. E. G. Peek\footnotemark[10]$^,$\footnotemark[11],
    Chad Schafer\footnotemark[12]$^,$\footnotemark[19],
    Padma A. Yanamandra-Fisher\footnotemark[13]$^,$\footnotemark[15],
    C.\,Alex Young\footnotemark[14]$^,$\footnotemark[15]

\medskip

\begin{raggedright}
\footnotetext[1]{\small{Cornell University, Cornell Center for Astrophysics and Planetary Science (CCAPS) \& Department of Statistical Sciences,  Ithaca, NY 14853, USA}}
\footnotetext[2]{\small{Division of Physics, Mathematics, \& Astronomy, California Institute of Technology,  Pasadena, CA 91125, USA}}
\footnotetext[3]{\small{Center for Astrophysics $|$ Harvard \& Smithsonian, Cambridge, MA 02138, USA}}
\footnotetext[4]{\small{eScience Institute, University of Washington, Seattle, WA 98195, USA}}
\footnotetext[5]{\small {DIRAC Institute, Department of Astronomy, University of Washington, Seattle, WA 98195, USA}}
\footnotetext[6]{\small{Department of Astronomy, University of Washington, Seattle, WA 98195, USA}}
\footnotetext[7]{\small{Penn State University, University Park, PA 16802, USA}}
\footnotetext[8]{\small{Booz Allen Hamilton, Annapolis Junction, MD, USA}}
\footnotetext[9]{\small{Department of Statistics \& Data Science, Yale University, New Haven, CT 06511, USA}}
\footnotetext[10]{\small{Department of Physics \& Astronomy, Johns Hopkins University, Baltimore, MD 21218, USA}}
\footnotetext[11]{\small{Space Telescope Science Institute, Baltimore, MD 21218, USA}}
\footnotetext[12]{\small {Department of Statistics \& Data Science
Carnegie Mellon University, Pittsburgh, PA, USA}}
\footnotetext[13]{Founder, The PACA Project, Space Science Institute, Boulder, CO 80301, USA}
\footnotetext[14]{NASA Goddard Space Flight Center, Greenbelt, MD 20771
USA}
\footnotetext[15]{American Astronomical Society Working Group on Astroinformatics and Astrostatistics}
\footnotetext[16]{American Astronomical Society Working Group on Time-Domain Astronomy}
\footnotetext[17]{American Statistical Association Astrostatistics Interest Group}
\footnotetext[18]{International Astronomical Union Commission B3 on Astroinformatics \& Astrostatistics}
\footnotetext[19]{LSST's Informatics and Statistics Science Collaboration (ISSC)}
\footnotetext[20]{International AstroInformatics Association}
\end{raggedright}

\pagebreak







\newpage
\pagenumbering{gobble}






\newpage
\pagenumbering{arabic}
\section{The growing impact of astrostatistics and astroinformatics}

Astrostatistics and astroinformatics (\ana) comprise interdisciplinary research combining astronomy with one or more of the information sciences, including statistics, machine learning, data mining, computer science, information engineering, and related fields.
For the Astro2010 decadal survey, nearly 100 astronomers and information scientists submitted two State of the Profession Position Papers \citep{borne2009,loredo2009} highlighting the potential of the then-emerging areas of astrostatistics and astroinformatics to make transformative contributions to astronomy, if only support for research and education in those areas could be enhanced.
In the decade since, the size and impact of \ana\ has grown dramatically, despite only modest changes in formal support of these areas.



In the time since Astro2010, the community of \ana\ researchers has grown tremendously in size.
Scholarly societies and large astronomy projects have responded with the creation of several \ana\ groups, with a combined membership of several hundred astronomers and information scientists: 
LSST's Informatics and Statistics Science Collaboration (ISSC, 2009, 72 members),
the International Astrostatistics Association\footnote{\url{http://iaa.mi.oa-brera.inaf.it/IAA/home.html}} (IAA, 2012, 601 members), 
the American Astronomical Society Working Group in Astroinformatics and Astrostatistics\footnote{\url{https://aas.org/comms/working-group-astroinformatics-and-astrostatistics-wgaa}}, (WGAA; 2012, 116 members)
the American Astronomical Society Working Group on Time Domain Astronomy\footnote{\url{https://aas.org/comms/working-group-time-domain-astronomy-wgtda}} (2014), 
the American Statistical Association Astrostatistics
Interest Group\footnote{\url{https://community.amstat.org/astrostats/home}} (2014, 111 members), 
the IEEE Astrominer Task Force (2014),
the International Astronomical Union Commission B3 on Astroinformatics \& Astrostatistics\footnote{\url{https://www.iau.org/science/scientific_bodies/commissions/B3/}} (2015, 239 members), 
and the International
AstroInformatics Association\footnote{\url{http://astroinformatics.info/}} (2019, 182 members).

Various teams from within the \ana\ community have submitted multiple Science White Papers addressing recent and future \ana\ science impacts in various areas of astronomy, and APC White Papers addressing specific \ana\  considerations such as the needs of petascale \ana\ research, and education and collaboration support issues.

This White Paper is authored by leaders of the \ana\ groups listed above, and reflects broad \ana\ support considerations discussed across their memberships. It briefly {\bf highlights the strong and growing impact of \ana, identifies key issues hampering the growth of this new field,
and offers recommendations for improved support of both research and education in \ana.}
This WP is not comprehensive; it does not address a number of astroinformatics issues, especially in the arenas of data systems, cyberinfrastructure, Virtual Observatory, etc..


At the turn of the century, SDSS \Dash the first large-scale, public, digital sky survey \Dash dramatically increased interest in statistics as well as machine learning and other computational sciences.
Indeed, SDSS is cited as an early example of the so-called Fourth Paradigm of science \Dash \emph{data-intensive science}, colloquially called ``big data science'' \citep{hey2009,Bell12009}.
Many astronomical data sources embody the classic ``three Vs'' \Dash {\textbf{\emph{volume, variety and velocity}}} \Dash distinguishing data-intensive science, particularly with recent wide-field surveys, optical/infrared integral field units, and radio interferometric instruments from earlier smaller and narrowly focused astronomy problems.
Time domain survey astronomy has emerged as a major endeavor as wide-field telescopes, both large and small, are dedicated to repeated photometric measurements of celestial populations, producing particularly large datasets.
Extracting sound science from complex data often requires advanced statistical and computational methods, even at relatively smaller volumes.
Challenging \ana\ research problems arise across the full spectrum of dataset scales. To highlight broader data science challenges, researchers have added other ``Vs'' to the list of big-data Vs, most notably \emph{veracity}, referring to the need to quantify uncertainty in data-based inferences, whether based on big datasets or modest ones.\footnote{\url{https://mapr.com/blog/top-10-big-data-challenges-serious-look-10-big-data-vs/}}
\begin{wrapfigure}{r}{0.41\textwidth}
\vspace{-0.4in}
\begin{center}
\includegraphics[width=0.45\textwidth]{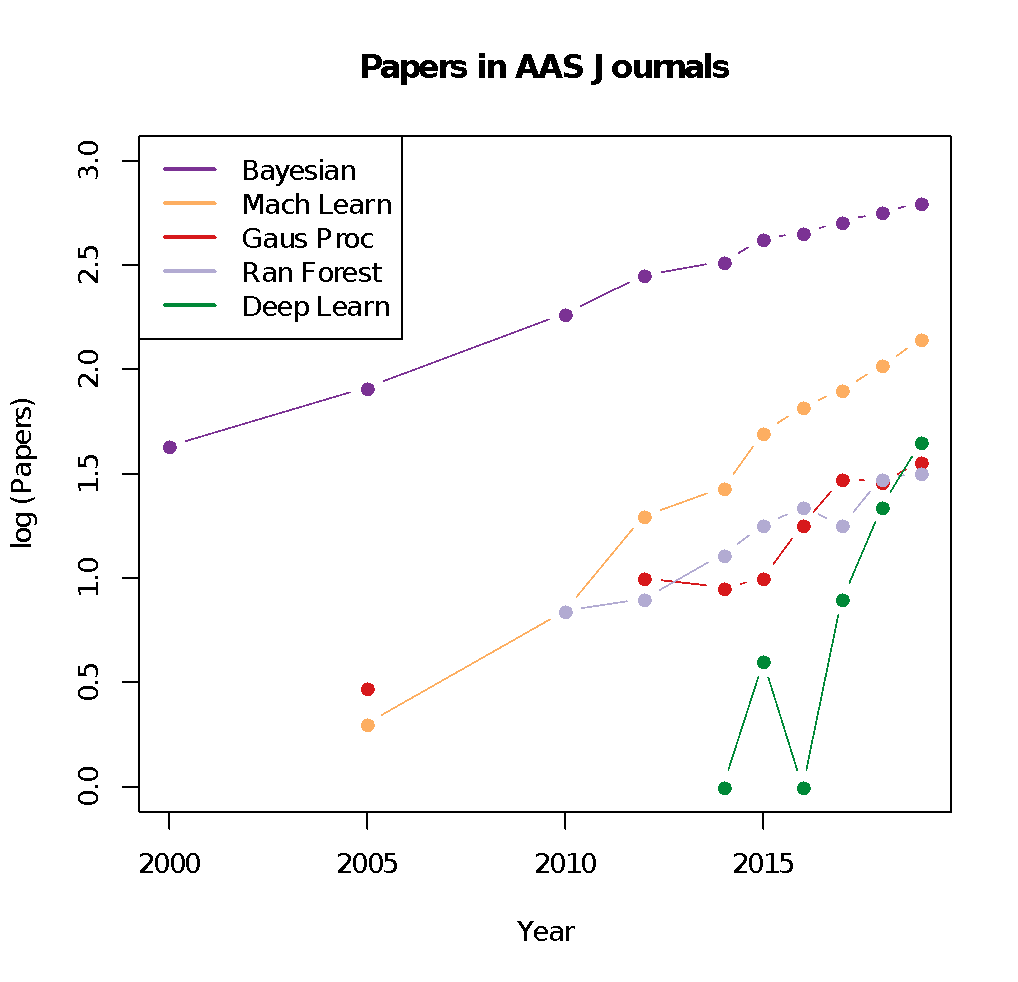}
\caption{\small Research papers using the emerging methodology
published in AAS Journals since 2000. The number of papers/yr is log10 scale. Methodology is  marked in colors: Bayesian - purple, Machine Learning - yellow, Gaussian Processes - red, Random Forests - grey, Deep Learning - green.  
}
\vspace{-0.4in}
\label{fig:pubs}
\end{center}
\end{wrapfigure}

Recent 
\emph{data challenges} provide excellent examples of the value of considering diverse methodologies for complex problems, and highlight the need to seek interdisciplinary collaboration. A decade ago, the Gravitational LEnsing Accuracy Testing (GREAT) weak lensing shear measurement competitions, GREAT08 and GREAT10 (including galaxy and star/PSF shape measurement challenges, \citealt{bridle2010,kitching2012,kitching2013}), were announced in \emph{Annals of Applied Statistics}. They drew submissions from dozens of teams, many submitting results from multiple methods, with several teams comprising non-astronomers. The GREAT08 prize went to a pair of computer scientists; the GREAT10 galaxy prize went to a pair of astronomers new to weak lensing, with organizers describing it as {\it ``a major success in its effort to generate new ideas and attract new people into the field.''} The more recent 2018 Photometric LSST Astronomical Time-Series Classification Challenge (PLAsTiCC, \citealt{kessler2019}) took advantage of newer crowd-sourcing tools (via the Kaggle platform), and attracted over 1000 submissions. Among the top five
performing teams, only a single participant was an astronomer.

The rise of advanced methodologies is recent and incredibly rapid with significant response by the research community.
Basic bibliometric statistics, displayed in Fig.~\ref{fig:pubs}, show that the use of many modern approaches and methods \Dash e.g., Bayesian statistics, machine learning, Gaussian processes, random forests, and deep learning \Dash is growing exponentially.
The \emph{Appendix} describes selected themes of emerging, advanced \ana\ research, highlighting the breadth of methods and applications, and the rapid growth of interest in adapting state-of-the art data science methods to astronomy.
During the 2013--19 period, the number of jobs emphasizing data analysis methodology offered to Ph.D. astronomers (both post-doc and faculty positions) approximately
doubled\footnote{\url{https://asaip.psu.edu/resources/jobs}}.  
The Astrostatistics Facebook group\footnote{\url{https://www.facebook.com/groups/astro.r/}} has over 4000 members with new members joining daily.
The community interest reflects the need for new methodology, and also for communication and training, as the standard training of astronomers lags behind.


The growth of interest in \ana\ research stands in contrast to serious deficiencies in support of the \ana\ education and research enterprise.
The following section highlights three key gaps between needs and available resources for realizing the potential of \ana\ to meet current and emerging science challenges.
The final section offers specific recommendations to close these gaps.

\section{Key Issues: Education, funding, and quality control}\label{sec:issues}
    
Recent developments in methodology were not widely anticipated and have proceeded rapidly.
This has resulted in unfamiliar challenges and imbalances for the educational, funding, and quality control structures of the field.  In the following subsections, we describe in detail the challenges and imbalances in each of these three areas.
From here, it is clear that actions are needed by different segments of the community \Dash universities, observatories and institutes, funding agencies, and leadership organizations like the National Academy of Sciences \Dash and we outline some recommendations in Section~\ref{sec:plan}.

\subsection{The education gap}
    
Astronomers are well-trained in mathematics relating to physical processes in order to do astrophysics, but not in applied mathematics, statistics, and computer science relating to extraction of reliable information from complex, noisy datasets. They are typically conversant with computer programming and processing on a moderate scale, but many are not prepared for the world of Big Data with challenges in data storage, access, and efficient analysis on high performance multicore computers, and modern software development practices.

The problem arises in the curriculum of physical scientists: courses in modern statistics, applied mathematics, and computer science are not in the required curriculum. For computation, this deficiency has been recently documented among astronomers: a survey of $\approx 1100$ astronomers found that 90\% write software but only 8\% received substantial training in software development \citep{momcheva2015}. Informal on-the-job training is adequate for some purposes, but
limits reproducibility, results in inefficient duplication, and
can lead to mediocrity, or even unnecessary failure, for more challenging problems. 
The methodology needed for astronomy and astrophysics is so diverse that specialized coursework in the usage of statistical software environments and computer resources is essential for the
astronomical research enterprise.

This deficit in the education of astronomers has been repeatedly noted: recent NASA\footnote{Big Data @ STScI: Enhancing STScI’s Astronomical Data Science Capabilities over the Next Five Years (2016), \url{http://archive.stsci.edu/reports/BigDataSDTReport\_Final.pdf}} and National Academy of Sciences\footnote{Optimizing the U.S. Ground-Based Optical and Infrared Astronomy System (2015), National Academy Press \url{https://www.nap.edu/}}
reports emphasize the need for professional training beyond long-standing formal education in the physical sciences. Some progress has been made. Textbooks on statistical methodology and data analysis (with computer codes) for astrophysics are available and taught in some universities \citep[e.g.,][]{feigelson_babu_2012,astroMLText,bailer-jones_2017,hilbe2017bayesian}\footnote{see \url{ https://asaip.psu.edu/resources/recent-books/methodology-books-for-astronomy}}. Informal Summer Schools, Hack Days, and tutorials have proliferated. However, except perhaps for disorganized Facebook-based discussion forums,
these resources are touching relatively few astronomers (perhaps 10\%).

University curricula are not renovating fast enough to match the
needs of methodological education for future space scientists, and
professional development resources are insufficiently funded or
organized to meet the needs of the research community.

\subsection{The funding gap}

Grants to universities and other 
institutions specifically designed to improve methodology for astronomical research are very scarce. NASA closed its only grant program in this area in 2011 Applied Information Science Research Program (AISRP). This program was critical, for example, to the development of the worldwide Virtual Observatory \citep{szalay2014}, and funded many smaller-scale \ana\ efforts, including development of SAOImage-DS9 and work on new statistical and machine learning algorithms by individual investigators and multi-university collaborations (the NASA \emph{Astrophysics Research, Analysis \& Enabling Technology 2011 Review Panel} evaluated the AISRP program in more detail\footnote{See the July 2011 section of the \href{https://science.nasa.gov/science-committee/subcommittees/nac-astrophysics-subcommittee}{NAC Astrophysics Subcommittee site}.}).
NSF has had short-lived interdisciplinary grant programs to promote mathematical developments for astronomy, and has supported astrostatistics at SAMSI programs. While astronomers do have access to agency-wide programs in cyberscience such as the Computational and Data-Enabled Science and Engineering (CDS\&E) and various cyber infrastructure programs, the success rate of such proposals may be {\em de facto} limited by level of buy-in from NSF's Division of Astronomical Sciences.

Other scientific fields do not have these structural problems. Biostatistics is taught in most universities and has been heavily funded by NIH for decades with many large grant programs\footnote{For illustration, the following NIH grant programs are available in cancer research (\url{statfund.cancer.gov/funding}): Big Data to Knowledge; Development of Informatics Technology; Informatics Technology for Cancer Research; Bridging the Gap between Cancer Mechanism and Population Science; Spatial Uncertainty: Data, Modeling, and Communication; Cancer Intervention and Surveillance Modeling Network; Short Courses on Mathematical, Statistical, and Computation Tools for Studying Biological Systems; NIDCR Grants for Data Analysis and Statistical Methodology applied to Genome-wide Data; NLM Express Analysis in Biomedical Informatics; Integrative Omics Data Analysis for Biomedical Informatics; New Computational Methods for Understanding the Functional Role of DNA Variants. 
}.  Statistics and informatics for Earth sciences
have been well-funded by the NSF and NASA\footnote{The following NSF grant programs are available in geosciences, in addition to agency-wide programs in mathematics and cyberscience: Collaboration in Mathematical Geosciences; EarthCube; Geoinformatics; Signals in the Soil; Advanced Digitization of Biodiversity Collections.  NASA operates the: Earth Observing System Data and Information System; NASA Center for Climate Simulation}, coordinated by the inter-agency Big Earth Data Initiative, with results presented in several dozen poster sessions at annual AGU meetings.

\subsection{The quality gap}

The culture of our research community and funding agencies is fully cognizant that major advances are driven by improvements in instrumentation, and that these instruments require software for operation and knowledge extraction. It is less well
recognized that new instruments give rise to science questions so diverse and complex that traditional data analysis procedures are often inadequate. The knowledge and skills of the statistician, applied mathematician, and algorithmic computer scientist need to be incorporated into programs that currently emphasize engineering and physical science in order to fully achieve the scientific potential of instruments and telescopes and the data they provide.  These issues might be divided into two stages: (1) data reduction through software pipelines (often developed within instrumentation groups, but ripe for methodological improvements from the broader community); and (2) science analysis that is performed by hundreds of scientists dispersed through U.S. universities and abroad. Both stages benefit from modern statistical and computational methods; in some cases, the science result is completely inaccessible without state-of-the-art methodology.

High standards for analysis methodology are not set consistently for publications, instrument analysis pipelines,
science analysis software developed by national observatories or space mission science centers, or for software produced by extramural science programs. The result is uneven quality in data and science analysis products; the methods used in astronomical software systems for data processing and science analysis are often inappropriate and/or obsolete 
\citep[e.g., see][]{protassov2002,tak2018}.
Limited peer review resources often make these kinds of problems undetectable until after publication.

\section{Strategic Plan}\label{sec:plan}


We have outlined an unusual situation for our profession:
the historical unfamiliarity of research based on advanced
cross-disciplinary methodology, and the rapidity of its growth, have led to imbalances that hinder research. Strides made in methodologies and computer science
are often not incorporated into astronomical research because we lack adequate educational, funding, and quality control structures. In the last decade or so, \emph{data science} has gained traction in both industry and academia; privately funded data science centers have appeared in industry and data science institutes have appeared in universities. While these centers and institutes have contributed to the development of new methodologies in A\&A, they are often not fully utilized by astronomy departments at universities and do not provide enough focused support for astronomy. Thus, a serious organizational commitment from the astronomy community at many levels is needed.

The problems outlined in Section~\ref{sec:issues} can be substantially rectified if concerted effort is made by the funding agencies, national observatories and mission centers, universities, and scholarly societies.
Large projects could not only fund software pipelines, but also cross-disciplinary study and oversight so the pipelines and associated science analysis software incorporate modern
statistical and computational methods. National
institutes could nurture internal teams devoted to methodology and hire consultants to advise large hardware and research groups. Universities can offer undergraduate and graduate courses in statistics, informatics, and the data sciences within astronomy programs, assuring students interested in data-intensive research careers sufficient curricular flexibility to become appropriately trained.


Cross-disciplinary interest groups that have emerged in scholarly societies can be energized with funds to
organize collaborative research efforts, conferences
and workshops, and informal education tutorials.

The obvious result of the lack of investment and commitment by the American astronomical enterprise in astrostatistics and astroinformatics is the loss of astronomical results, particularly relating to Big Data science from LSST and its predecessor instruments. During the past decade, \ana\ has been established as an important research area in the astronomical community. The importance of this research should be also recognized by agencies and universities, and supported by appropriate changes in the funding and educational structures.

With these issues in mind, we offer the following recommendations.
We estimate that the total new cost for implementing our specific research and training recommendations is a few million dollars annually, a small fraction of annual spending in astronomy.
This small investment will have a disproportionately large impact on A\&A and on astronomy as a whole.
Our team does not have the resources and expertise to assess costs in detail; further, several recommendations involve adjusting the balance of various resources (monetary and otherwise) across multiple stakeholders.
\emph{We propose that the Astro2020 survey recommend that the AAS or another appropriate body establish a committee to review the support of \ana, using these recommendations as a starting point.}
The committee should be provided sufficient resources and access to stakeholders to enable developing detailed and realistic recommendations for improved support of \ana.


\subsection{Closing the education gap}

\subsubsection{Universities and National Observatories}

\begin{itemize}
    \item Universities should revise the curriculum in undergraduate physical science to require courses in applied statistics, mathematics, and computer science. At the graduate level, specialized courses in computational methods and usage of statistical methods should be incorporated into the astronomy and astrophysics curriculum. Specifically, students should learn how to use modern astronomy computing environments, and how to harness modern computing hardware efficiently.
    
    \item Universities and national observatories should develop information science courses for astronomers at the undergraduate and graduate levels.
    
    \item Universities and national observatories should financially support summer schools and cross-disciplinary workshops on advanced methods, both to train astronomical data science researchers and to integrate this emerging area into mainstream astronomy.
    
    \item Universities and national observatories should establish specialized permanent appointments for data science in astronomy, as routinely as they now do for observers/instrumentalists and theorists. Cross-appointment permanent positions (e.g., with statistics departments, computer science departments, etc.) should also be considered.  
    
    \end{itemize}
    
    \subsubsection{NSF and NASA}
    
    \begin{itemize}

    \item NSF and NASA should establish mechanisms to support educators interested in developing, curating, improving, maintaining, and/or disseminating astroinformatics materials that accelerate and improve astroinformatics education in the community.

    \item NSF and NASA should survey their existing programs, at the agency and center levels (e.g. NASA centers), which support the A\&A education of the community within these agencies. This includes programs such as the Frontier Development Lab from NASA Ames, NASA Goddard’s astropy summer schools, and the astroinformatics working groups at Goddard (and other centers). 
\end{itemize}

\subsection{Closing the funding gap}

\subsubsection{Universities and National Observatories}

\begin{itemize}
    \item University astronomy departments and National Observatories should work with internal data science institutes and other departments (e.g. statistics, mathematics, computer science) to offer competitive, interdisciplinary postdoctoral fellowships in \ana.
    
    \item  Universities should financially support multidisciplinary PhDs in astronomy (e.g., in \ana). This would also encourage graduates to enter astronomy programs even if they are interested in possibly pursuing a more general career in data science.
    
\end{itemize}

\subsubsection{NSF and NASA}

\begin{itemize}

    \item NSF should provide interdisciplinary grant support for research related to \ana. 
    
    \item NSF Division of Astronomical Sciences should pursue partnerships in support of medium and large projects that have a significant astronomical data science component.

\item With community input, NASA should be urged to reorganize its support of data analysis and information science research. There should be a focus on financial support for both routine and advanced data analysis research that serves space-based astrophysics through development, adaptation, validation and application of modern \ana\ methods.

\item Both NASA and NSF \ana\ research programs should implement explicitly multi-tiered support, with different categories of research of various duration and levels of funding. Long-term funding must be included, especially targeting young researchers.  

\item NASA and NSF should encourage reviewers of  postdoctoral fellowship applications to recognize \ana, including both proposed \ana\ research and/or a track-record of high-quality \ana\ in previous research publications. 

\item NASA and NSF should instate a 3-year interdisciplinary fellowship program in astronomical data sciences. This would encourage young scientists to pursue \ana\ careers, and would bring recognition to these scientists and to the discipline.  

\item NASA and NSF should also support astronomical data science research targeting \textbf{infrastructure}
(e.g., data management and computational resource management research, including development of
astronomy-oriented parallel, grid, and cloud computing software
environments, and maintaining the critical software tools).  
Such support should be separated from support from focused, \emph{science-driven
data science research}, either via separate programs, or via explicitly
identified proposal categories within a single program.

\item Similarly, NASA and NSF should support the development of significant public, open-source software that provides important science-enabling technology, much like the development of a new instrument.
As with instruments, significant codebases need maintenance, and funding channels need to support major updates of widely-used codebases similar to how instrument maintenance is supported.

\item Agencies should develop or adapt funding opportunities enabling support of \ana\ data challenges, like those mentioned in \S~1. Data challenges (e.g. PLAsTiCC) draw in participants from communities outside of astronomy, and could lead to more interdisciplinary collaboration and higher quality research.

\item We echo the recommendations made in the NASA Task Force on Big Data for SMD\footnote{\url{https://science.nasa.gov/science-committee/subcommittees/big-data-task-force}}: 
 
{\it Recommendation:  SMD should 
establish a new division that would focus on cross-cutting data science and computing projects and whose responsibilities would include establishing the Data Science Applications Program which will promote bringing modern data science methodologies into SMD’s data analysis
worlds including the science operations of SMD’s missions.

Recommendation:  
In staffing the Science Committee and the four thematic Science Advisory Committees, SMD should ensure that at least one appointment on each of these committees is reserved for an expert who is a routine user of high-performance computers (NASA’s or others), is active in employing modern data science methodologies, and/or is deeply involved in the science operations of large, complex scientific data archives. 

Recommendation: NASA should make prioritized investments in computing and analysis hardware, workflow software and education and training to substantially accelerate modeling workflows. NASA should take the lead to make substantial increases in:  ... software modernization; resources to develop new data analysis paradigms; education and training workshops, scientific conferences and journal special collections to effect a culture acceptance of the importance of workflow development and management; … lossy data compression and more advanced methods for signal detection.

}


\end{itemize}

\subsection{Closing the quality gap}

\begin{itemize}
    \item Journals should maintain high standards for analysis methodology and algorithms. This may involve supporting a statistics/informatics editor (as is currently done by the AAS journals), and modifying review processes to ensure that papers with significant \ana\ content are examined by reviewers with expertise in both the relevant astrophysics and the relevant information science.
    Authors should be strongly encouraged (perhaps required, in some circumstances)
    to make computational results reproducible, e.g., by publishing software, repositories, and/or
    computational ``notebooks'' along with papers.
    
    \item Astronomy curricula should include training in best practices in development of software, e.g., by encouraging or requiring formal training in programming and development practices from faculty actively engaged in astroinformatics education, computer science departments, and/or from well-vetted training programs (e.g., Software Carpentry).
    
    \item Support interdisciplinary collaborations that seek funding to include substantive contributions from experts in specific algorithms and/or computational methods that could advance their research goals.

\item Funding agencies should encourage production of open-source software; this development model improves code quality in broad scientific applications.


\end{itemize}

\section{Appendix: Emerging data science themes in astronomy}


Here we briefly survey a selection of important data science areas where recent developments in statistics and machine learning are beginning to make significant impacts in astronomy\footnote{see also the Science White Paper submitted to the Astro2020 in March 2019 \citep{siemiginowska2019}}.
This survey is by no means exhaustive, nor are the highlighted applications meant to be endorsements of specific approaches.
Rather, this survey is intended to display the broad scope of data science research in astronomy and its potential for producing qualitative advances in our ability to distill science from data.
Few if any of the highlighted approaches are covered in the recent spate of books on statistics and machine learning in astronomy, emphasizing the need for interdisciplinary research and collaboration.

\textbf{Nonlinear dimensionality reduction.}
Many astronomical datasets ``live'' in a high-dimensional space.
An observed spectral energy distribution comprising measurements in dozens, hundreds, or thousands of spectral bands may be considered to be a vector in a sample space with a dimension for each band.
An image with a million spatial pixels may be considered as a vector in a million-dimensional sample space.
Empirically, a collection of many cases of such data very often lies on or near a low-dimensional manifold in the full sample space.
Discovering such a manifold can enable dramatic improvement in inference tasks such as classification or regression (characterizing correlations).
When that manifold is a hyperplane, it may be found using techniques from linear algebra that are well-known to astronomers, \emph{principal component analysis} (PCA) being the best-known example.
But more often, the manifold will be a complex curve or surface, and discovering it requires tools for \emph{nonlinear dimensionality reduction}.
This has been a major research area in statistics and machine learning in the last decade, with several new techniques making significant impacts in diverse areas of astronomy.
Several of the most successful approaches rely on analysis of the matrix of pairwise distances or similarities of the data (\emph{spectral clustering} or \emph{spectral connectivity analysis}).
Examples include use of diffusion maps for supernova classification, locally-biased spectral graph analysis for describing SDSS galaxy spectra, t-Distributed Stochastic Neighbor Embedding (t-SNE) for classification of supernovae and stellar spectra, and convolutional neural network based autoencoders for radio galaxy classification \citep{richards2011,lawlor2016,lochner2016,reis2018, ma19}.






\textbf{Sparsity.}
A similar but complementary kind of reduction in complexity can occur in the parameter space used to describe the ``true'' signals underlying observed data. 
Signals are often best described in a \emph{transform} space, e.g., Fourier space for periodic time-domain signals, or wavelet or shapelet spaces for images.
Empirically, many natural signals have very sparse representations in an appropriately selected transform space.
E.g., although the Fourier transform of a light curve with N time samples has O(N) Fourier coefficients, periodic light curves can be described with many fewer than O(N) coefficients.
Similarly, image compression exploits the observation that natural images with N pixels are well-described with many fewer than O(N) coefficients in, say, a discrete cosine transform (DCT) or wavelet basis.
Information scientists are devising new models and algorithms that exploit knowledge of sparsity to improve signal recovery.
A notable example is \emph{compressed sensing}, a class of signal processing techniques{~ }that exploits sparsity in a transform space to enable recovery of complex signals even when the \emph{data} are relatively sparse (e.g., not fully covering Fourier space).
Example applications in astronomy include improving image recovery in radio interferometry, measuring cosmological image distortions due to weak lensing, and inverting solar flare differential emission measure (DEM) data.
The method now appears in about 30 astronomy papers/yr. \citep{hastie2015,wiaux2009,leonard2012,carrillo2014,cheung2015}

\textbf{Deep learning.}
Many flexible model architectures in statistics and machine learning are built by composition of a large number of simple elements.
Examples include basis expansions (e.g., linear combinations of Fourier or wavelet basis functions) and artificial neural nets (ANNs, linear combinations of simple but nonlinearly tunable ``activation functions'').
Early theoretical work on the approximation power of ANNs showed that the flexibility of such compositions could be greatly enhanced by layering: for models with fixed ``width'' (the number of linearly combined components), flexibility can be greatly enhanced by adding ``depth'' (using the outputs of components of one layer as inputs to a new layer of superposed components). 
Since the early 2000s, advances in training algorithms for deep models, combined with wide availability of massively parallel computing capability via GPUs and large training sets, have enabled deep learning (DL) algorithms to leapfrog competitors in many industrial applications (e.g., speech and image recognition and classification).
A key aspect of these algorithms has been inclusion of dimension-reducing layers, e.g., via tunable convolution or subsampling operations.
These enable DL models to discover rich \emph{hierarchical feature representations} of data, e.g., describing an image in terms of edges with different orientations at the lowest level, groups of edges comprising more complex features at the next level, and so on.
It has long been appreciated that feature selection is crucial to the performance of machine learning algorithms; DL models can partly automate feature selection.
DL is being applied to diverse learning tasks in many areas of astronomy, e.g., classification of galaxy images and stellar spectra, image deblending, photometric redshift estimation, classification of variables and transients, and source discovery in multimessenger astronomy \citep{Goodfellow2016,hoyle2016,Mahabal2017,pasquet2019,allen2019,boucaud2019,wu2019,muthukrishna2019}.










\newpage
\bibliographystyle{natbib}
\bibliography{astro2020.bib}

\end{document}